\documentclass[11pt]{article}
\usepackage{graphicx} 
\usepackage{amsmath}
\usepackage{amssymb}
\usepackage{amsthm}
\usepackage{tabularx}
\usepackage{graphicx} 
\usepackage[margin=1in]{geometry}
\usepackage{hyperref}
\usepackage{xcolor}
\usepackage{multicol}
\usepackage{url}
\usepackage[utf8]{inputenc}
\usepackage{enumitem}
\usepackage{fontawesome}

\usepackage{tikz}
\usetikzlibrary{shapes.geometric}
\usetikzlibrary{arrows.meta}
\usetikzlibrary{calc}

\usepackage{todonotes}

\newcommand{\polylog}{\operatorname{polylog}}
\newcommand{\Ham}{\operatorname{Ham}}

\newcommand{\Span}{\operatorname{span}}

\newcommand{\Enc}{\operatorname{Enc}}

\newcommand{\rank}{\operatorname{rank}}

\newcommand{\poly}{\operatorname{poly}}

\newcommand{\N}{\mathbb{N}}
\newcommand{\F}{\mathbb{F}}
\newcommand{\FF}{\mathbb{F}}

\newcommand{\eps}{\varepsilon}

\newcommand{\E}{\mathbb E}

\newcommand{\R}{\mathbb{R}}
\newcommand{\Z}{\mathbb{Z}}

\usepackage{aliascnt}
\usepackage{cleveref}

\newtheorem{theorem}{Theorem}

\newaliascnt{lemma}{theorem}
\newtheorem{lemma}[lemma]{Lemma}
\aliascntresetthe{lemma}

\newaliascnt{claim}{theorem}

\aliascntresetthe{claim}

\newaliascnt{proposition}{theorem}

\aliascntresetthe{proposition}

\newaliascnt{corollary}{theorem}
\newtheorem{corollary}[corollary]{Corollary}
\aliascntresetthe{corollary}

\newaliascnt{question}{theorem}
\newtheorem{question}[question]{Question}
\aliascntresetthe{question}

\newaliascnt{conjecture}{theorem}

\aliascntresetthe{conjecture}

\theoremstyle{definition}

\newaliascnt{definition}{theorem}
\newtheorem{definition}[definition]{Definition}
\aliascntresetthe{definition}

\newaliascnt{remark}{theorem}
\newtheorem{remark}[remark]{Remark}
\aliascntresetthe{remark}

\newaliascnt{example}{theorem}

\aliascntresetthe{example}
\usepackage{graphicx} 

\usepackage{titlesec}
\titlespacing*{\paragraph}{0pt}{1ex plus 0.5ex minus 0.2ex}{1em}
\title{\textbf{Explicit Subspace Evasive Sets via Subspace Designs}}
\author{Joshua Brakensiek\thanks{University of California, Berkeley, CA, USA. Email: \href{mailto:josh.brakensiek@berkeley.edu}{\texttt{josh.brakensiek@berkeley.edu}}. Research supported in part by Venkatesan Gurusawmi's Simons Investigator award and NSF grants CCF-2211972 and DMS-2503280.} \and Yeyuan Chen\thanks{University of Michigan, Ann Arbor, MI, USA. Email: \href{mailto:yeyuanch@umich.edu}{\texttt{yeyuanch@umich.edu}}. Research partially supported by the National 
Science Foundation under Grant No.\ CCF-2236931.} 
\and Mahdi Cheraghchi\thanks{University of Michigan, Ann Arbor, MI, USA. Email: \href{mailto:mahdich@umich.edu}{\texttt{mahdich@umich.edu}}. Research partially supported by the National 
Science Foundation under Grant No.\ CCF-2236931.} 
\and Rohan Goyal\thanks{Massachusetts Institute of Technology, Cambridge, MA, USA. Email: \href{mailto:rohan_g@mit.edu}{\texttt{rohan\_g@mit.edu}}.  Research supported by Yael Tauman Kalai's NSF Award 6956054.} \and Venkatesan Guruswami\thanks{Department of Electrical Engineering and Computer Sciences, Department of Mathematics, and the Simons Institute for the Theory of Computing, UC Berkeley, Berkeley, CA 94709, USA. Email: \href{mailto:venkatg@berkeley.edu}{\texttt{venkatg@berkeley.edu}}. Research supported in part by a Simons Investigator Award, a UC Noyce initiative award, DOD Advanced
Research Projects Agency grant HR0011262E031, and NSF grant 2211972.} \and Aaron Putterman\thanks{Harvard University, Cambridge, MA, USA. Supported in part by a Jane Street Graduate Research Fellowship, the Simons Investigator Awards of Madhu Sudan and Salil Vadhan, and AFOSR award FA9550-25-1-0112. Email: \href{mailto:aputterman@g.harvard.edu}{\texttt{aputterman@g.harvard.edu}}.} \and Zihan Zhang \thanks{Simons Institute for the Theory of Computing, University of California Berkeley, Berkeley, CA, USA. Email: \href{mailto:zihanz@hku.hk}{\texttt{zihanz@hku.hk}}.}}
\date{October 2026}
\begin{document}

\maketitle
\thispagestyle{empty}
\begin{abstract}
Given a finite field $\mathbb F_q$, we say that a set $A\subseteq \mathbb F^n_q$ is an $(r,c)$ subspace evasive set if and only if for any affine subspace $U\subseteq \mathbb F^n_q$ with bounded dimension $\dim(U)\leq r$, the intersection size $|A\cap U|$ is at most $c$. The probabilistic method shows the existence of $(r,c)$ subspace evasive sets with $|A| = q^{(1-\epsilon) n}$ with $c = O(r/\epsilon)$. However, the best explicit constructions by Dvir--Lovett (STOC 2012) and Ben-Aroya--Shinkar (CJTCS 2013) have $c$ growing as an exponential function of $r$.

Here, we give an explicit construction of such subspace evasive set that achieves the intersection bound $c=O(\min(r\log(r/\epsilon)/\epsilon, r^2/\epsilon))$ when both $r$ and $1/\eps$ are constants. Our approach is simple and the key idea is to compose subspace design codes (Guruswami-Xing, STOC 2013) with a suitably chosen ``local'' subspace evasive set. As a corollary, we give the first construction of an explicit (strongly) capacity-achieving list-recoverable code with nearly optimal list size, circumventing all previously-known barriers to construction.
\end{abstract}

\newpage
\section{Introduction}



In this work, we study \emph{explicit} constructions of subspace evasive sets, a fundamental object with connections to pseudorandomness, coding theory, and Ramsey theory. Formally, given a finite field $\F_q$, we say that a set $A\subseteq \F^n_q$ is an $(r,c)$ subspace evasive set if and only if for any affine subspace $U\subseteq \F^n_q$ with bounded dimension $\dim(U)\leq r$, the intersection size $|A\cap U|$ is at most $c$. Intuitively, the set $A$ should have minimal overlap with any low-dimensional affine space. As a classical example, the case $r=1$ corresponds to finding a collection of points in $\F_q^n$ such that no $c+1$ points lie on the same line; the case $r=2$ corresponds to no $c+1$ points lying in the same plane, and so forth. 

This notion was first introduced by Pudl{\'a}k and R{\"o}dl~\cite{pr04}, motivated by an application to bipartite Ramsey graphs. Specifically, they showed that an explicit construction of an $(n/2,c)$ subspace evasive set $A\subseteq \F^n_2$ can be used to construct explicit $c$-Ramsey bipartite graphs with $|A|$ vertices (where a $c$-Ramsey bipartite graph is a bipartite graph with no empty or complete $c$ by $c$ subgraph). Using this connection, they gave the first construction of $o(\sqrt{N})$-Ramsey bipartite graphs on $N$ vertices.\footnote{The problem has since seen major progress with other approaches and we now know explicit $(\log N)^{O(1)}$-Ramsey bipartite graphs; still a near-optimal explicit subspace-evasive set in $\F_2^n$ would be a very pleasing way to construct $\Theta(\log N)$-Ramsey bipartite graphs. \cite{cz16,li23}}

Inspired by this initial line of work, many subsequent works extended the study of subspace evasive sets and found concrete applications of subspace evasive sets to other areas in theoretical computer science. For instance, the work of Guruswami~\cite{guru11} showed that explicit constructions of subspace evasive sets with good quality lead to explicit list decodable/recoverable codes. Subsequently, Dvir and Lovett \cite{dl12} gave explicit subspace evasive sets which when plugged into \cite{guru11,GuruswamiW2013} gave capacity-achieving list-decodable codes with $O(1)$ list-size. These subspace evasive sets also played a crucial role in the design of better affine extractors \cite{zewi2011affine} and even in combinatorics \cite{ST24}. In this work, we \emph{reverse} some of these established connections, and show that recent work in designing better list-recoverable codes actually leads to much-improved explicit subspace evasive sets.

To make our discussion more concrete, let us fix $0<\eps<1$ and $r\ge 1$. Our goal in this paper is to design explicit $(r,c)$ subspace evasive sets $A\subseteq \F^n_q$ with size at least $|A|\ge q^{(1-\eps)n}$, while ensuring that $c$ is as small as possible. As a simple baseline, one can consider what happens when we select $A$ \emph{uniformly at random}.
By a simple probabilistic argument (see, for instance, Lemma 8 of \cite{guru11}), if one samples $A$ of size $q^{(1-\eps)n}$ in this way, one can show that it will be a $(r,O(r/\eps))$ subspace evasive set with high probability. 
Complementing this, the work of Sudakov and Tomon~\cite[Theorem 1.1]{ST24} shows that there is a matching lower bound. Indeed, any $A$ with size $q^{(1-\eps)n}$ \emph{cannot} be an $(r,o(r/\eps))$ subspace evasive set when $\eps<1/20$ and $r=(1+\Omega(1))\log_2{(1/\eps)}$. Thus, for most regimes, the \emph{optimal} $(r,c)$-subspace evasive sets have $c = O(r/\eps)$.

The central question in the study of subspace evasive sets is the pursuit of \emph{explicit} constructions, where one can algorithmically determine the elements of our subspace evasive set $A\subseteq \F^n_q$ efficiently. Motivated by the broad applications in \cite{pr04,guru11}, and the hope that better explicit subspace evasive sets can lead to downstream explicit constructions of other combinatorial objects we pose the following challenge:
\begin{question}\label{mainques}Can we provide explicit constructions of $(r,c(r,\eps))$ subspace evasive set $A\subseteq \F^n_q$ with size at least $|A|\ge q^{(1-\eps)n}$, ideally matching the probabilistic method bound of $c(r,\eps)=O(r/\eps)$?
\end{question}

In the special case $r=1$, F{\"u}hrer and Taranchuk~\cite{lineses} construct a $(1,O(\sqrt{1/\eps}))$ subspace evasive set, which is actually better than what can be guaranteed by the probabilistic method. In the general case, the pioneering work of Dvir and Lovett~\cite{dl12} gives explicit constructions with $c(r,\eps)=(\frac{r}{\eps})^{r}$, and follow-up work due to Ben-Aroya and Shinkar~\cite{note13} refines this bound to $c(r,\eps)=(\frac{2}{\eps})^{r}$ when $r$ and $\eps$ are both constants.

For general $r$, there has been little progress on Question~\ref{mainques} in the last decade; however, we invite the reader to \Cref{subsec:related-work} for discussion of many related works. As mentioned above, the main contribution of our work is to give explicit constructions of \emph{near-optimal} $(r,c(r,\eps))$ subspace evasive sets.

\begin{theorem}[Main Theorem, informal, see \Cref{sec:explicit-SES}]\label{thm:main-informal}
For any constants $r \ge 1$, $\eps > 0$, for all large enough $n$, one can explicitly construct an $(r, c(r,\eps))$ subspace evasive set $A\subseteq \F^n_q$ with size $|A|\ge q^{(1-\eps)n}$ and intersection size $c(r,\eps)=O(\min(r^2/\eps, r\log(r/\eps)/\eps))$.
\end{theorem}

Our bound on $c(r,\eps)$ exponentially improves upon the previous work and is much closer to the probabilistic method bound of $c(r,\eps)=O(r/\eps)$. In turn, by using the reductions of \cite{guru11} and its follow-ups in conjunction with our new constructions, we resolve some open problems in coding theory. We elaborate upon these applications and more of our techniques in the following section.

\subsection{Our Results and Techniques}\label{subsec:our-results}

For simplicity of presentation, we slightly modify the definition of a subspace evasive set. In particular, we replace the condition on affine sets (as stated in \cite{dl12,lineses}) with subspaces.


\begin{definition}[Subspace evasive set \cite{pr04,dl12,GuruswamiW2013}]\label{def:SES} Let $\F$ be a field, a set $A\subseteq \F^n$ is $(r,c)$ subspace evasive if for any subspace $U\subseteq \F^n$ with dimension $\dim(U)\leq r$, we have $|U\cap A|\leq c$.
\end{definition}
Equivalently, $A$ is $(r,c)$ subspace evasive set if every subset of $A$ of size more than $c$ has rank more than $r$. This change of definition is overall mild, as any $r$-dimensional affine space is contained in an $r+1$-dimensional linear space, so an $(r+1,c)$ subspace evasive set \Cref{def:SES} captures the more traditional notion of an $(r, c)$ subspace evasive set. In particular, our asymptotic bounds--such as \Cref{mainques}--hold in either setting.


At the heart of our results is the notion of \emph{list recovery}. In short, a code $C \subseteq \Sigma^n$ (not necessarily linear) is $(\ell, L)$ zero-error list-recoverable if for any sets $S_1, \hdots, S_n \subseteq \Sigma$ of size at most $\ell$, the set $C \cap (S_1 \times \cdots \times S_n)$ has size at most $L$. Previously, Guruswami~\cite{guru11} had shown that explicit subspace evasive sets can be used to construct explicit list-recoverable codes with bounded output list-size. Our primary technical contribution is a \emph{constructive converse} that explicit list-recoverable codes in combination with a small ``local'' subspace evasive set can be used to explicitly construct a large ``global'' subspace evasive sets. 

This relies on a modified notion of list-recoverability which we call \emph{subspace list-recoverability}:


\begin{definition}[Subspace list-recoverability]\label{def:slrc}
For a local dimension $s$, and a global dimension $n$, we say that a code $C \subseteq (\F^s)^n$ is $(r, \rho, \ell, L)$ subspace list-recoverable if the following holds: for every $\F$-linear subspace $U \subseteq (\F^s)^n$ with $\dim(U) \leq r$, and every choice of $S_1, \dots S_n \in \binom{\F^s}{\leq \ell}$, we have 
\[
\left | \{c \in C \cap U  \,\colon\, \exists y \in (S_1 \times \cdots \times S_n), \Ham(c, y) \le \rho n\} \right | \leq L.
\]
Here $\Ham(c, y)=\sum^n_{i=1}[c_i\neq y_i]$ denotes the Hamming distance.
\end{definition}

Intuitively, this is a \emph{weaker} requirement than being a standard $(\rho, \ell, L)$ list-recoverable code, which requires that for any $S_1, \dots S_n \in \binom{\F^s}{\leq \ell}$,
\[
\left | \{c \in C  \,\colon\, \exists y \in (S_1 \times \cdots \times S_n), \Ham(c, y) \le \rho n\} \right | \leq L.
\]
Our new definition of \emph{subspace list-recoverability} only requires that after taking an \emph{intersection} with a low-dimensional linear space (albeit an arbitrary one), the resulting code satisfies the ordinary list-recovery guarantee. In particular, any $(\rho, \ell, L)$ list-recoverable code is also $(r, \rho, \ell, L)$ list-recoverable for every $r$. As we shall see, this distinction is particularly useful when constructing subspace evasive sets, as we must only ever bound the number of points that lie in the intersection with a fixed-dimensional subspace. 


\paragraph{Key Idea: Composition.} The core conceptual insight of this paper is that the \emph{composition} of a subspace list-recoverable code with a ``local'' subspace evasive set is sufficient to construct a ``global'' subspace evasive set with near-optimal parameters. This composition does need a (mild) structural requirement that the subspace evasive sets are graphs of functions. We encapsulate this observation as follows.

\begin{theorem}[Informal version of \Cref{thm:framework}]\label{thm:framework-informal}
Let $\mathrm{Enc} : \F^k \to (\F^s)^n$ be the encoding map of an $\F$-additive\footnote{$\F$-additivity means that for any $c, c' \in C$, and $\alpha,\beta \in \F$, $\alpha c + \beta c' \in C$ as well. In other words, $C$ is an $\F$-linear space.} code $C \subseteq (\F^s)^n$ which is $(r, 0, c, L)$ list-recoverable. Let $f : \F^s \to\F^b$ be a function such that $A = \{(a, f(a)) \;:\; a \in \F^s\} \subseteq \F^{s+b}$ is an $(r, c)$ subspace evasive set. 
Then, their composition
\[
    B = \{(x, f(\mathrm{Enc}(x)_1), \hdots, f(\mathrm{Enc}(x)_n)) \;:\; x \in \F^k\} \subseteq \F^{k+bn}
\]
is an $(r, L)$ subspace evasive set of size $q^k$. 
\end{theorem}

If we think of the subspace evasive set itself as a code, this construction has close parallels to \emph{code concatenation} in coding theory. Intuitively, \Cref{thm:framework-informal} shows that one can use subspace list-recovery to effect a local-to-global principle that \emph{amplifies} a small subspace evasive set into a larger one.

To explain concretely why this is the case, consider any $r$-dimensional subspace $U \subseteq \F^{k + bn}$, and consider all messages $X \subseteq \F^k$ whose corresponding points $B_X := \{(x, f(\mathrm{Enc}(x)_1), \hdots, f(\mathrm{Enc}(x)_n)) : x \in X\}$ land in $U$. Since $X$ is the projection of $B_X$ onto its first $k$ coordinates, we have that $X$ is itself contained in an $r$-dimensional subspace. The challenge still remains to show that $|X|$ is not too large.

Proving this relies on the \emph{local} subspace evasive set. For every $i \in [n]$, we consider the following projection of $B_X$:
\[
    A_{X,i} := \{(\mathrm{Enc}(x)_i, f(\mathrm{Enc}(x)_i)) : x \in X\} \subseteq A.
\]
Since $\mathrm{Enc}()$ is $\F$-linear, $A_{X,i}$ is a linear image of $B_X$, and it is contained in an $r$-dimensional space. Furthermore, since $A_{X,i}$ is contained in an $(r, c)$ subspace evasive set, we have that $A_{X,i}$ and thus $S_i := \mathrm{Enc}(X)_i$ has cardinality at most $c$. Therefore, $\Enc(X) \subseteq S_1 \times \cdots \times S_n$, where each $S_i$ has cardinality at most $c$. On the other hand,  $X$ is contained in an $r$-dimensional subspace. Since $\Enc$ is the encoding map of an $(r, 0, c, L)$ subspace list-recoverable code, we can conclude $|X| \le L$. Therefore, $B$ is an $(r, L)$ subspace evasive set.

Taking a step back, it should not be so surprising that this combination of subspace evasive sets and list-recoverable codes is fruitful. List-recoverable codes guarantee that in any set with enough codewords, there must be some coordinate where the codewords take many different values. Subspace-evasive sets guarantee that if any set contains enough points, then those points span a high dimensional space. The composition in \Cref{thm:framework-informal} uses the first property to find a coordinate with many different values, and the second property to ensure those values have high rank. This property of list-recoverable codes has also found many other surprising uses, including recent oracle separations of quantum complexity classes (see \cite{YamakawaZ2024, bostanci2026separating}).

\paragraph{The Building Blocks.} Thanks to \Cref{thm:framework-informal}, we have reduced the task of constructing an explicit subspace evasive set into two problems: (1) efficient construction of the ``local'' subspace evasive set and (2) construction of a good additive subspace list-recoverable code. For the former challenge, we can efficiently derandomize the probabilistic construction of subspace evasive sets using the method of conditional expectations. In short, for a ``semi-random'' set $A$ where some elements are deterministic and some are randomly chosen, we efficiently estimate the expected number of violations to our subspace evasive set property and derandomize one-at-a-time the elements of $A$, while keeping the expected number of violations monotonically decreasing.

For the latter challenge of constructing good subspace list-recoverable codes, we can  use many recent advances in this space off-the-shelf. For example, the recent work of Brakensiek, Chen, Dhar, and Zhang~\cite{BrakensiekCDZ2025-BrascampLieb} (see also the work of Goyal and Guruswami~\cite{GoyalG2025-Structure}) showed that any \emph{subspace design code} can be used to construct very good list-recoverable codes. 
Subspace designs, introduced by Guruswami and Xing~\cite{GuruswamiX2013}, are a pseudorandom family of linear maps  which have found many applications including list-decoding (their original motivation in \cite{GuruswamiX2013} and recently in \cite{Srivastava2025,GoyalG2025-Proximity}), list recovery~\cite{BrakensiekCDZ2025-BrascampLieb,GoyalG2025-Structure}, rank condensers and dimension expanders~\cite{ForbesG2014,GRX21}, and algorithmic derandomization~\cite{BrakensiekCDZ2025-Explicit,chatterjee2026bipartite}, and proximity gaps of codes ~\cite{GoyalG2025-Proximity,GoyalGSW2026}. Many of these applications have been discovered quite recently and have thus motivated a detailed investigation of better constructions of subspace designable codes~\cite{GuruswamiK2016,GoyalGH2026,GuoRSZ2026,cohen2026algebraic,AlrabiahGG26a}. As such, by composing these subspace design constructions with \cite{BrakensiekCDZ2025-BrascampLieb,GoyalG2025-Structure}, we get a number of explicit subspace evasive set constructions achieving \Cref{thm:main-informal}, see \Cref{sec:explicit-SES} for exact statements.


\paragraph{Implications for List-recovery.}  

\begin{table}[ht]
\footnotesize
\begin{tabular}{|m{7cm}|m{3cm}|m{3cm}|m{2cm}|m{1.4cm}|}
\hline 
Codes & Alphabet size $|\Sigma|$ & List size $L$ & Explicit?    \\ \hline

Random Linear Codes \cite{ZP81,guruthesis,rw18,ls25} & $\ell^{O(1/\eps)}$ & $(\ell/\eps)^{O(\ell/\eps)}$ & No\\
\hline
 Random Linear Codes \cite{BrakensiekCDZ2025-BrascampLieb,gg26b} & $(\frac{\ell}{R+\eps})^{O((R+\eps)/\eps^2)}$ & $\left(\frac{\ell}{R+\epsilon}\right)^{O(1+R/\eps)}$ & No\\
\hline
Random Codes & $\ell^{O(1/\eps)}$ & $O(\ell/\eps)$ & No\\
\hline
 \hline 
 Folded RS/Univariate Multiplicity Codes \cite{GuruswamiR2008,kopparty2015list,GuruswamiW2013,kopparty2018improved,BrakensiekCDZ2025-BrascampLieb}& $\left( \frac{ n \ell}{\eps^2} \right)^{O(\ell/\eps^2)}$ &$\left(\frac{\ell}{R+\epsilon}\right)^{O(1+R/\eps)}$&Yes\\
\hline
Folded RS Codes \& Subspace Evasive Sets \cite{dl12} & $\left( \frac{n}{\eps^2} \right)^{O(\log(\ell)/\eps^2)}$& $(\frac{1}{\eps})^{O(1/\eps^3)}$ & Yes    \\ \hline 
Codes Based on the AEL Framework \cite{SrivastavaT2025,JeronimoS2025,GoyalGH2026}& $\exp(\exp(O(\ell^2/\eps^2)))$ & $\left(\frac{\ell}{R+\epsilon}\right)^{O(1+R/\eps)}$ &Yes \\ \hline
Folded RS plus AEL Codes \cite{kopparty2018improved} & $(\frac{\ell}{\eps})^{O(\ell^2\log{(\ell/\eps)}/\eps^2)}$& $\ell^{O(1/\eps^4)}$ & Yes     \\ \hline 
Folded AG Codes \cite{GuruswamiX2013,GuruswamiK2016} & $\exp\left( \frac{ \ell \log(\ell/\eps) }{\eps^2} \right)$ & $2^{2^{2^{O_{\eps,\ell}(\log^*(n))}}}$ & Yes  \\ \hline
AG Codes + BTT evasive subspace \cite{grz21} & $(\frac{\ell}{\eps})^{O(\ell/\eps^2)}$ & $\exp(\poly(\ell/\eps))$ & Yes  \\ \hline
\hline
RS codes + Subspace Evasive Sets \textbf{(\Cref{thm:subrs-main-res})}& $n^{O(1/\eps^2)}$ & $O(\frac{\ell}{\eps^3}\log{\frac{\ell}{\eps}})$ &Yes \\
\hline
AG codes + Subspace Evasive Sets \textbf{(\Cref{thm:subag-main})}& $(\frac{\ell}{\eps})^{O(\ell/\eps^2)}$ & $\poly(\ell/\eps)$ &Yes \\
\hline
Folded RS codes + Subspace Evasive Sets \textbf{(\Cref{thm:subfrs-main-res})} & $(\frac{n\ell}{\eps^2})^{O(\ell/\eps^2)}$ & $O(\frac{\ell}{\eps^2}\log{\frac{\ell}{\eps}})$ &Yes \\
\hline
AEL codes + Subspace Evasive Sets \textbf{(\Cref{thm:subael-main-res})}& $\exp(\exp(\ell^2/\eps^2))$ & $O(\frac{\ell}{\eps^2}\log{\frac{\ell}{\eps}})$ &Yes \\
 \hline
 \hline
 Any Linear Code \cite{guru21,ChenZ2025,LeviMS2025,ls25} & Any & $\ell^{\Omega(R/\eps)}$ & Lower Bound\\
 \hline
 Classical AEL Codes \cite{bcpz26} & Any & $\ell^{\Omega(R/\eps)}$ & Lower Bound\\
 \hline
\end{tabular}
\caption{This table gives a summary of many constructions of list-recoverable codes as well as some known limitations.} \label{tab:listrecovery}
\end{table}

As observed by Guruswami~\cite{guru11}, explicit subspace evasive sets can be used to construct explicit list-recoverable codes. In broad strokes, we prove a family of results of the following form.
\begin{theorem}[Informal, see \Cref{sec:explicit-LR,sec:frs-alg}]
For any constants $\ell\ge 1,\eps,R\in(0,1)$, there are explicit codes with rate at least $R$ that are $(1-R-\eps,\ell, \poly(\ell/\eps))$ list-recoverable. Furthermore, for the construction stated in \Cref{thm:alg-frs-main}, both the encoding and the list-recovery of the code can be done in $O_{R,\ell, \eps}(1) \poly(n)$ time.
\end{theorem}
See \Cref{tab:listrecovery} for a comparison of our results with prior work, the most notable contrast being that the list size of all previous constructions had exponential dependence in $1/\eps$. (See \Cref{tab:listrecovery-app} for a more comprehensive comparison.) Some of our constructions get a list size of $O(\frac{\ell}{\eps^2}\log{\frac{\ell}{\eps}})$, which is not too far off the probabilistic method guarantee of $O(\ell/\eps)$.


At first glance, it may seem counterintuitive that our construction has novel implications for list-recoverable codes, as state-of-the-art constructions of list-recoverable codes are a key input to \Cref{thm:framework-informal}. However, note that \Cref{thm:framework-informal} requires as input an \emph{$\F$-additive} code, but a subspace evasive set by definition has no additive structure. Thus, any list-recoverable code constructed using our subspace evasive sets circumvents lower bounds on the list-recoverability of linear and additive codes~\cite{guru21,ChenZ2025,LeviMS2025,ls25}. Furthermore, this technique for constructing list-recoverable codes falls outside of the recently-popularized ``AEL framework''~\cite{ael95} for constructing list-decodable and list-recoverable codes~\cite{SrivastavaT2025,JeronimoS2025,GoyalGH2026}, thus circumventing recently-established lower bounds in that space as well~\cite{bcpz26}.

All of our constructions work by taking an established code construction (such as Reed--Solomon codes) and judiciously using the subspace evasive set to select a subcode with good list-recovery guarantees. For example, the work of Guruswami and Xing \cite{gx22} (see also {\cite[Section 5.3]{ldsurvey}) shows that\footnote{These works only discuss the list-decoding case where $\ell=1$, but it is straight-forward to generalize the whole proof to general list-recovery for any $\ell\ge 1$. See \cite{gz26} for details.} there are explicit additive subcodes of Reed-Solomon codes such that the lists of codewords obtained in the list-recovery problem lie in a low-dimensional $O(\ell/\eps^2)$-dimensional subspace of the message space. By taking a further subcode of this Reed--Solomon subcode based on our subspace evasive set construction within its message space,  we get that this $O(\ell/\eps^2)$-dimensional subspace has a small intersection with our subcoding as a subspace evasive set, yielding the desired $\poly(\ell/\eps)$ list-size bound. We obtain analogous constructions for algebraic-geometry codes (based on \cite{grz21}), folded Reed--Solomon codes (based on \cite{GuruswamiW2013}), and AEL codes (based on \cite{GoyalGH2026}). Each of these constructions yield different tradeoffs between alphabet size, list size, and encoding/decoding time (some of which have linear-dependence in $n$). See \Cref{sec:explicit-LR} for detailed statements of all of these constructions. 








\subsection{Related Work}\label{subsec:related-work}



We now give some broader context to subspace evasive sets by briefly discussing generalizations and connections in the literature. See in particular the recent paper of Lim, Nie and Zeng~\cite{LimNZ26} which surveys a number of these works.

\paragraph{Variety-evasive Sets.}
A natural generalization of $(r, c)$ subspace evasive sets is requiring intersection of size at most $c$ for families of non-affine sets $V \subseteq \F_q^n$, such as solutions of polynomial equations of higher degree. In particular, we say that a set $A \subseteq \F_q^n$ is $(d, r, c)$ variety evasive if for every variety $V \subseteq \F_q^n$ of degree at most $d$ and dimension at most $r$, we have that $|V \cap A| \le c$. The case $d = 1$ coincides with the notion of subspace evasiveness. Variety evasive sets were introduced by Dvir, Koll{\'a}r and Lovett~\cite{DKL14} as a direct generalization of Dvir and Lovett's~\cite{dl12} investigation of subspace evasive sets.  Variety evasive sets should not be confused with the variety evasive subspace families \cite{Guo24,AndrewsGarg26} which also are motivated by applications to derandomization.

\paragraph{Subspace-evasive Subspaces.}
In the study of dimension expanders, Guruswami, Resch, and Xing~\cite{GRX21} define a notion of \emph{subspace evasive subspaces}. Here, one constructs an $\F_q$-subspace $H \subseteq \F_q^n$ which has small intersection with every $\F_{q^d}$-subspace $W \subseteq \F_{q^d}^{n/d}$ which can also be viewed as an $\F_q$-subspace $W \subseteq \F_q^n$.

\paragraph{Twisted Varieties.} Closely related to variety evasive sets are the concept of \emph{twisted} varieties, where one imposes that $A \subseteq \F_q^n$ is an $n-r$-dimensional variety, where we impose that for any degree-$d$, $r$-dimensional variety $V \subseteq \F_q^n$, the intersection $V \cap A$ is $0$-dimensional. This notion was introduced by works of Dvir--Koll{\'a}r--Lovett~\cite{dl12,DKL14} and Blagojevi{\'c}--Bukh--Karasev~\cite{BBK13}, the latter of whom called these ``non-degenerated embeddings.'' See the work of Lim, Nie and Zeng~\cite{LimNZ26} for new bounds on the existence of twisted varieties.

\paragraph{Connections to Extremal Combinatorics.} Evading varieties has recently found much interest~\cite{IhringerV22,SZ26,LimNZ26} due to the concept being a natural generalization of ``points in general position.'' In particular, studying variety-evasive sets (even existentially) has a number of applications to incidence geometry~\cite{ST24,MTS24} and hypergraph Tur{\'a}n problems~\cite{BBK13,CLNZ25}. As mentioned previously, in certain parameter regimes, subspace evasive sets can also be used to construct explicit Ramsey graphs~\cite{pr04}.

\paragraph{Applications of List-Recoverable Codes}
List-recoverable codes themselves have found many important applications in the literature. For instance, list-recoverability plays an important role in the construction of unbalanced lossless expanders \cite{GuruswamiUV2009} and in oracle separations of quantum complexity classes \cite{YamakawaZ2024, bostanci2026separating}.


\subsubsection{Paper Structure}
In \Cref{sec:l2g}, we define subspace list-recoverable codes, and give the general framework  to construct the ``global'' subspace evasive sets given explicit subspace list-recoverable codes (SLRCs) and ``local'' subspace evasive sets. In \Cref{sec:prior}, we list explicit constructions of useful SLRCs as well as describe a construction of ``local'' subspace evasive sets using the method of conditional expectation. In \Cref{sec:explicit-SES}, we plug these prior constructions into the framework and get explicit subspace evasive sets with required qualities. In \Cref{sec:explicit-LR}, we use the constructed subspace evasive sets to construct good list-recoverable codes as shown in \Cref{tab:listrecovery}. In the end, we show in \Cref{sec:algorithm} how to speed up the list-recovery of such codes using known techniques from coding theory.

\paragraph{Notation.} For any set $A$ and non-negative integer $t\ge 0$, $\binom{A}{t}$ denote the collection of subsets of $A$ with size $t$, and $\binom{A}{\leq t}$ denote the collection of subsets of $A$ with size at most $t$. For any vector $x\in \Sigma^n$, let $x_{\leq m}\in\Sigma^m$ denote the length-$m$ prefix of $x$ and $x_i\in\Sigma$ denote the $i$-th coordinate of $x$. For a set of vectors $X\subseteq \Sigma^n$, let $X_{\leq m}$ denote $X_{\leq m}=\{x_{\leq m}\colon x\in X\}$ and $X_{m}$ denote $X_{m}=\{x_{m}\colon x\in X\}$. We also define $x_{\ge m}$ and $X_{\ge m}$ similarly.

Given an $\F$-additive code $C \subseteq (\F^s)^n$, an encoder of $C$ is an $\F$-linear function $f\colon \F^k\to(\F^s)^n$ with $\mathsf{Im}(f)=C$. If some encoder $f$ can be computed in time $T$, we say $C$ is encodable in time $T$.

For any subset $A \subseteq \F^n$, we say that $A$ is constructable in time\footnote{We assume all field operations take time $O(1)$.} $T$ if there is a bijection $f\colon \{1, 2, \hdots, |A|\} \to A$ which can be computed in time $T$, including any preprocessing time needed to determine the function $f$. In a similar vein, we say that $A$ has a membership oracle in time $T$ if the indicator function $g \colon \F^n \to \{0,1\}$ defined by $g(x) = \mathbf{1}[x \in A]$ can be computed in time $T$.

\section{Local to Global Amplification via Subspace List-recoverable Codes}\label{sec:l2g}

In this section, we show how to use subspace list-recoverable codes (SLRCs) to amplify local subspace evasive sets to global subspace evasive sets. Recall from Definition~\ref{def:slrc} that a code $C \subseteq (\F^s)^n$ is $(r, \rho, \ell, L)$ subspace list-recoverable if for any product set $S_1 \times \cdots \times S_n \subseteq (\F^s)^n$, where each $S_i$ has cardinality at most $\ell$, any $r$-dimensional subcode of $C$ has at most $L$ codewords within relative Hamming distance $\rho$ of this product set. In particular, subspace list-recovery is a weaker notion than traditional list-recovery, and an $\F$-additive code $C\subseteq (\F^s)^n$ that is $(\rho,\ell,L)$ list-recoverable is automatically $(r,\rho,\ell,L)$ subspace list-recoverable for any $r \ge 0$.


This local-to-global reduction actually requires the ``local'' subspace evasive set to have a \emph{generator function}, which is defined below.

\begin{definition}[Generator function]
Let $f\colon \F^a\to\F^b$ be a function, we define $G^f\colon=\{(x,f(x))\colon x\in\F^a\}\subseteq \F^{a+b}$ to be the set generated by $f$ (i.e., the \emph{graph} of $f$), and we say $f$ is the generator function of $G^f$.
\end{definition}

We now prove our core theorem, which shows how to construct a ``global'' subspace evasive set from a ``local'' subspace evasive set and a subspace list-recoverable code.
\begin{theorem}\label{thm:framework}
Let $\F$ be a field, $f\colon \F^s\to \F^b$ be the generator function of an $(r,c_f)$ subspace evasive set $G^f$, and $C\subseteq (\F^s)^n$ with $\dim_{\F}(C)=k$ be an $\F_q$-additive $(r,0,c_f,L)$ subspace list-recoverable code. We can construct a function $g\colon \F^k\to \F^{bn}$ such that $B=G^g=\{(x,g(x))\colon x\in \F^k_q\}\subseteq \F^{k+bn}$ is an $(r,L)$ subspace evasive set with size $|B|=q^k$.

Moreover, let $f$ can be computed in time $T_f$ and $C$ is encodable in time $T_C$, then $g$ can be computed in time $O(T_C+nT_f)$, $B$ is constructable in time $O(T_C+nT_f)$, and $B$ has membership oracle in time $O(T_C+nT_f)$.
\end{theorem}
\begin{proof}
Let $\Enc \colon \F^k\to(\F^s)^n$ denote an $\F$-linear encoder of $C$ so that $\mathsf{Im}(\Enc)=C$ and $\Enc$ can be computed in time $T_C$. We construct $g\colon\F^k\to \F^{bn}$ as follows:
\[
g(x) := (f(\Enc(x)_1),\dots,f(\Enc(x)_n))\in\F^{bn},\ \ \forall x\in\F^k.
\]
We first prove that $B=\{(x,g(x))\colon x\in \F^k_q\}\subseteq \F^{k+bn}$ is indeed an $(r,L)$ subspace evasive set. Let $U\subseteq \F^{k+bn}$ denote any subspace with $\dim(U)\leq r$. Let $U'=U_{\leq k}\subseteq \F^k$ denote the projection of $U$ on the first $k$ coordinates. Let $A=U\cap B$ and $A'=A_{\leq k}$ similarly. Since $A\subseteq B$, by the construction of $B$ we know distinct elements $x,y$ in $A$ have distinct length-$k$ prefixes $x_{\leq k},y_{\leq k}$. Therefore, $|A|=|A'|$. It suffices to prove $|A'|\leq L$.

Suppose for contradiction that $|A'| > L$. For each $i\in[n]$, let $\Enc_i\colon \F^k\to\F^s$ denote the linear map outputting the $i$'th coordinate of the encoding map $\Enc$ of $C$, i.e., $\Enc_i(x)=\Enc(x)_i$ for all $x\in\F^k$. Since $\Enc(A')\subseteq \Enc(U')$ with $\dim(\Enc(U'))\leq r$, and $C$ is $(r,0,c_f,L)$ subspace list-recoverable, we know because $|A'| > L$ that there must be some $i\in[n]$ so that $|\Enc_i(A')|> c_f$.
 
Now define $H_i =\{(\Enc_i(x),f(\Enc_i(x)))\,\colon\, x\in A'\}$. Note that $|H_i| \ge |\Enc_i(A')|> c_f$. Since $G^f \supseteq H_i$ is an $(r, c_f)$ subspace evasive set, we must have that $\rank (H_i) > r$. This implies that
\begin{align*}
    r \ge \dim(U) &\ge \rank (A) = \rank \{(x, g(x)) \,\colon\, x \in A'\}\\&\ge \rank \{(\Enc_i(x), f(\Enc_i(x))) \,\colon\, x \in A'\} = \rank (H_i) > r,
\end{align*}
a contradiction. Therefore, we must have that $|A| = |A'| \le L$. This proves that $B$ is an $(r,L)$ subspace evasive set.

Finally we justify the complexity aspects of our construction. Evaluating $g$ requires an evaluation of $\Enc$ and $n$ evaluations of $f$, and thus $O(T_C+nT_f)$ time. In terms of the construction,  we can compute a bijection $f_B\colon |B|\to B$ as $f_B(x)=(x,g(x))$ by one evaluation of $g$, which requires $O(T_C+nT_f)$ time.  To implement the membership oracle for the constructed subspace evasive set, for a given $(x\in\F^k,y\in\F^{bn})$, we need to compute $g(x)$ to see whether $g(x)=y$. This can be implemented in $O(T_C+nT_f)$ time.
\end{proof}

\section{Building Blocks}\label{sec:prior}


In this section, we discuss methods for constructing ``local'' subspace evasive sets, and record existing constructions of subspace list-recoverable codes from the literature.

\subsection{Local Subspace Evasive Sets via Random Sampling}

We first present how to construct the ``local'' subspace evasive set via pre-conditioning methods, which is the function $f$ that we use in \Cref{thm:framework}.
The following construction can be seen as an intelligent brute-force search. More concretely, it is a derandomization of the probabilistic method union bound, which can be seen as the toy case of the derandomization of the Lov\'{a}sz Local Lemma \cite{cgh09}. 

\begin{theorem}\label{thm:inner}
Let $s,b,r,d>0$ be integers and $\F_q$ be a finite field, if $(s+b+d)r<bd$, then we can construct, in time $O(q^{(s+b)(r+1)+rd}d)$, a function $f\colon \F^s_q\to\F^b_q$ so that $G^f\subseteq \F^{s+b}_q$ is an $(r,d-1)$ subspace evasive set.
\end{theorem}
\begin{proof}
Let $\F^s_q=\{x_1,\dots,x_{q^s}\}$ and  $g\colon \F^s_q\to\F^b_q$ denote a random function. There are at most $q^{(s+b)r}$ many $r$-dimensional subspaces $U$ of $\F^{s+b}_q$. For each of these subspaces $U$, it has at most $q^{rd}$ subsets $S\subseteq T$ with $|S|= d$. Therefore, there are at most $q^{(s+b+d)r}$ possible pairs $(S,T)$. For each of such $S\subseteq T$, consider the bad event $E_{S,T}$ when $S\subseteq G^g= \{(x,g(x))\in\F^{s+b}_q\colon x\in\F^s_q\}$ happens. Our goal is to construct a function $f\colon \F^s_q\to\F^b_q$ such that no $E_{S,T}$ happens conditioned on $f=g$.

We assign values to $f(x_1),\dots,f(x_{q^s})$ one by one. For each $i\in[q^s]$, let $I_i=\{x_1,\dots,x_{i}\}$ and $f_i\colon I_i\to\F^b_q$ denote the assigned values on $I_i$. Let $p_{i}=\E[\sum_{S,T}E_{S,T}\,\colon\, f_{i}=g|_{I_{i}}]$. We start from $i=0$ and $f_i=\emptyset$. 

For each $i\in[q^s]$, in the $i$-th round,  we have the following.
\[
p_{i-1}=\E\left[\sum_{S,T}E_{S,T}\,\colon\, f_{i-1}=g|_{I_{i-1}}\right]=\frac{1}{q^b}\sum_{w\in\F^b_q}\E\left[\sum_{S,T}E_{S,T}\,\colon\, f_{i-1}=g|_{I_{i-1}}\wedge g(x_i)=w\right]
\]
Computing these expectations requires $O(q^{b+(b+s+d)r}d)$ time. Therefore, one can choose some $w\in\F^b_q$ such that $\E[\sum_{S,T}E_{S,T}\,\colon\, f_{i-1}=g|_{I_{i-1}}\wedge g(x_i)=w]\leq p_{i-1}$ in deterministic $O(q^{b+(b+s+d)r}d)$ time. By assigning $f_i(x_i)=w$ and $f_i|_{I_{i-1}}=f_{i-1}$, we can guarantee that $p_{i}\leq p_{i-1}$ and proceed to round $i+1$.

This whole procedure requires $O(q^{(b+s)(r+1)+rd}d)$ time. We can guarantee that \[p_{q^s}=\E[\sum_{S,T}E_{S,T}\,\colon\, f=g]\leq p_0=\E[\sum_{S,T}E_{S,T}].\] Since we succeed when $p_{q^s}<1$, it suffices to show that $p_0=\E[\sum_{S,T}E_{S,T}]<1$.

Fix any valid $S\subseteq T$. Let $S=\{a_1,\dots,a_{|S|}\}$. $E_{S,T}$ happens iff for all $i\in[|S|]$, there is $g(a_{i,\leq s})=a_{i,\ge s+1}$. For a uniformly random function $g$, the probability that this happens is $q^{-b}$. 

If $|S|>|S_{\leq s}|$, then since $G^g$ is generated by $g$, it be the case that $E_{S,T}=0$. Otherwise, we have that \[\Pr[E_{S,T}]\leq q^{-b|S_{\leq s}|}\leq q^{-b|S|}\leq q^{-bd}.\]
Since $(s+b+d)r<bd$, we have that
\[
p_0=\E\left[\sum_{S,T}E_{S,T}\right]\leq \sum_{S,T}q^{-bd}\leq q^{(b+s+d)r-bd}<1,\]
as desired.
\end{proof}
\begin{remark}
From \cite[Lemma 23]{GuruswamiW2013}, there exists a polynomial $p\in\F_{q^s}[X]$ with degree at most $d$ such that $\{(x,p(x)_{\leq b})\in\F^{s+b}_q\colon x\in\F_{q^s}\}$ achieves the similar subspace evasiveness as \Cref{thm:inner}. We can enumerate all such polynomials $p(x)$ and find one with the target property. This gives an alternative and more space-efficient construction of the local subspace evasive sets we need.
\end{remark}
\subsection{Explicit Subspace List-recoverable Constructions}\label{sec:slcc}
We next list useful explicit constructions of subspace list-recoverable codes which follows from recent advances in pseudorandomness and coding theory.

We first define lossless rank-condenser and states its implications for SLRCs. We make use of this notion in \Cref{sec:cons3}.
\begin{definition}[\cite{Forbes2014}]\label{def:lossless-rank-condenser}
    An additive code $C\subseteq \left(\FF_q^s\right)^n$ forms a lossless rank-condenser up to rank $r$ if for any subspace $U\subseteq C$ with $\dim U\le r$, we have that there is an $i\in [n]$ such that $\pi_i: \left(\FF_q^s\right)^n \mapsto \FF_q^s$, the projection map onto the $i$th coordinate, is injective on $U$.
\end{definition}

\begin{lemma}\label{def:condeneser-to-SLRC}
    If an additive code $C\subseteq \left(\FF_q^s\right)^n$ is a lossless rank-condenser up to rank $r$, then it also forms a $(r, 0, \ell, \ell)$ subspace list-recoverable code for all $\ell\in \N$.
\end{lemma}

\begin{proof}
    Consider any $U\subseteq C\subseteq \left(\FF_q^s\right)^n$ with $\dim U\leq r$ as well as $S_1,\dots, S_n \in \binom{\FF_q^s}{\leq \ell}$. Define $\mathcal L =\{c \in U\cap C \,\colon\, \forall i\in [n], \pi_i(c)\in S_i\}$. Then there exists an $i\in [n]$ such that $\pi_{i}|_{U}$ is injective. Now,  $|\pi_i(\mathcal L)|=|\mathcal L|$ but $\pi_i(\mathcal L) \subseteq S_i$. Hence, $|\mathcal L|\leq |S_i|\leq \ell$.
\end{proof}

Next, we define subspace design codes \cite{GuruswamiX2013} and demonstrate their implications for SLRCs. We make use of such codes in \Cref{sec:cons1,sec:cons2}.

\begin{definition}[Subspace-Design Property \cite{GuruswamiX2013, ChenZ2025, GoyalG2025-Proximity}]\label{def: Subspace-designs}
    For any function $\tau: \mathbb N\rightarrow \R_{\le 1}$, an $\FF_q$-additive code $\mathcal C\subseteq (\FF_q^s)^n$ is said to be a $\tau$-\textit{subspace design} code if for every $r\in \mathbb N$, and every $\F_q$-linear subspace $\mathcal A$ of $\mathcal C$ of dimension at most $r$, the following holds: 
    \begin{equation*} \frac{1}{n}\sum\limits_{i=1}^n \dim \mathcal A_i\le \dim(\mathcal A)\cdot \tau(r) \end{equation*}
    where $\mathcal A_i = \{a\in \mathcal A\,\colon\, a_i=0\}$.
\end{definition}

\begin{lemma}\label{lem:SD-to-condenser}
    If an additive code $C\subseteq \left(\FF_q^s\right)^n$ is a $\tau$-subspace design with $\tau(r)<1/r$, then it also forms a $(r, 0, \ell, \ell)$ subspace list-recoverable code for all $\ell\in \N$.
\end{lemma}

\begin{proof}
    Consider any $U\subseteq C\subseteq \left(\FF_q^s\right)^n$ with $\dim U\leq r$ as well as $S_1,\dots, S_n \in \binom{\FF_q^s}{\leq \ell}$. Define $\mathcal L =\{c \in U\cap C\,\colon\, \forall i\in [n], \pi_i(c)\in S_i\}$. Observe that now, $\frac{1}{n}\sum_{i=1}^n \dim U_i<1$
    where $U_i = \{u\in U\,\colon\, u_i =0\}$. Thus, there exists an $i\in [n]$ with $\dim U_i = 0$. Hence $\pi_i$ is injective on $U$. Therefore by \Cref{def:condeneser-to-SLRC}, we have that $C$ is $(r,0,\ell,\ell)$ subspace list-recoverable. 
\end{proof}

\begin{theorem}[subspace list-recovery of subspace-designs \cite{BrakensiekCDZ2025-BrascampLieb, GoyalG2025-Structure}]\label{thm:SD-SLRC}
    If $\mathcal C\subseteq \Sigma^n$ is a $\tau$-subspace design code, then for any $\eps>0$, $C$ is $\left(r, 1-\tau(r)-\eps, \ell, \bigl(\tfrac{\ell}{\tau(r)+\epsilon}\bigr)^{(\tau(r)+\epsilon)/\epsilon}\right)$ subspace list-recoverable.
\end{theorem}

A proof of the above fact is found in \cite[Appendix A, Theorem A.4]{GoyalG2025-Structure}.

\begin{corollary}\label{cor:SD-0SLRC}
    If $\mathcal C\subseteq \Sigma^n$ is a $\tau$-subspace design code, then $C$ is $\left(r, 0, \ell, \ell^{1/(1-\tau(r))}\right)$ subspace list-recoverable. Additionally, when $\tau(r)\le 1/(1+\log_2\ell)$, then $C$ is $(r, 0, \ell, 2\ell)$  subspace list-recoverable.
\end{corollary}
\begin{proof}
    Set $\eps = 1-\tau(r)$ in \Cref{thm:SD-SLRC}. Setting $\tau(r)=\frac{1}{1+\log \ell}$ gives that $C$ is $(r, 0, \ell, \ell^{\frac{1+\log_2 \ell}{\log_2 \ell}})=(r, 0, \ell, 2\ell)$ subspace list-recoverable. 
\end{proof}

We now instantiate \Cref{lem:SD-to-condenser} and \Cref{cor:SD-0SLRC} with specific choices of subspace design codes.

\subsubsection{Construction 1: \cite{GoyalGH2026} subspace designs}\label{sec:cons1}

We now present the first construction of subspace list-recoverable codes via the subspace designs of Goyal, Guruswami, and Hsieh~\cite{GoyalGH2026} based on AEL codes~\cite{ael95}.

\begin{theorem}[\cite{GoyalGH2026}]\label{thm:ggh}
    For any finite field $\FF_q$, $R, \eps>0$ with $R+\eps\leq 1$, and dimension $r\in \mathbb N$, there exists an $s = O_{q, r, \eps}(1)=\poly(r,1/\eps)\cdot q^{r^2}$ such that for all sufficiently large $n$, there exists an explicit code $ C\subseteq(\FF_q^s)^n$ of rate $\ge R$,  that forms a $\tau$-subspace design with $\tau(r)\leq R+\eps$. In particular, $C$ can be computed in time $O_{q,r, \eps}(1)\cdot \poly(n)$.
\end{theorem}

By combining \Cref{thm:ggh} with \Cref{lem:SD-to-condenser}, we get a family of SLRCs with the following parameters.

\begin{corollary}\label{cor:ggh1}
    For any finite field $\FF_q$ and $r\in \mathbb N$, for all large enough $n$, there exists an explicit $\tau$ subspace-design additive code $C\subseteq (\FF_q^s)^n$ with rate $R \ge 1/2r$, $\tau(r)<1/r$, $s=O_{q,r}(1)=\poly(r)\cdot q^{r^2}$ which can be computed in time $O_{q,r}(1)\cdot \poly(n)$. In particular, this code is an $(r,0,\ell, \ell)$ subspace list-recoverable code for all $\ell \in \mathbb N$.
\end{corollary}

Likewise, combining \Cref{thm:ggh} with \Cref{cor:SD-0SLRC}, obtains the following family of SLRCs.

\begin{corollary}\label{cor:ggh2}
    For any finite field $\FF_q$ and $r, \ell\in \mathbb N$, for all large enough $n$, there exists an explicit $\tau$ subspace-design additive code $C\subseteq (\FF_q^s)^n$ with rate $R \ge \frac{1}{2+2\log_2\ell}$, $\tau(r)\leq \frac{1}{1+\log_2\ell}$, $s=O_{q,r}(1)=\poly(r,\log \ell)\cdot q^{r^2}$ which can be computed in time $O_{q,r}(1)\cdot \poly(n)$. In particular, this code is an $(r,0,\ell, 2\ell)$ subspace list-recoverable code for all $\ell \in \mathbb N$.
\end{corollary}

\subsubsection{Construction 2: \cite{GuruswamiK2016} subspace designs}\label{sec:cons2}

We next present a construction of subspace list-recoverable codes via the subspace designs of Guruswami and Kopparty~\cite{GuruswamiK2016}.

\begin{theorem}[\cite{GuruswamiK2016}]\label{thm:gk}
    For any $R \in (0,1)$ and $s, n\in \mathbb N$ such that $Rsn$ is an integer, and any finite field $\mathbb F_q$ with $q>sn$, there exists an explicit additive $\tau$-subspace design code with $\tau(r)\leq \frac{sR}{s-r+1}$ for all $r\leq s$. Additionally, the code can be computed in time $n\cdot \poly(s, \log q)$.\footnote{This code is in fact the \cite{GuruswamiR2008} folded Reed-Solomon code.}
\end{theorem}

Combining \Cref{thm:gk} with \Cref{lem:SD-to-condenser} yields the following SLRCs.

\begin{corollary}\label{cor:gk1}
    For any $r\in \mathbb N$, and integer $n\ge 2$, and finite field $\FF_q$ with $q>4nr$, there exists an explicit $\tau$ subspace-design additive code $C\subseteq (\FF_q^s)^n$ with rate $R \ge 1/8r$, $\tau(r)<1/4r$, $s=2r$ which can be computed in time $n\cdot \poly(r, \log q)$. In particular, this code is an $(r,0,\ell, \ell)$ subspace list-recoverable code for all $\ell \in \mathbb N$.
\end{corollary}

Furthermore, combining \Cref{thm:gk} with \Cref{cor:SD-0SLRC} obtains the following SLRCs.

\begin{corollary}\label{cor:gk2}
    For any $r,\ell \in \mathbb N$ with $n>4\log \ell$, and finite field $\FF_q$ with $q>4nr$, there exists an explicit $\tau$ subspace-design additive code $C\subseteq (\FF_q^s)^n$ with rate $R \ge 1/(3+3\log_2\ell)$, $\tau(r)<1/(1+\log_2\ell)$, $s=2r$ which can be computed in time $n\cdot \poly(r, \log q, \log \ell)$. In particular, this code is an $(r,0,\ell, 2\ell)$ subspace list-recoverable code.
\end{corollary}

\subsubsection{Construction 3: \cite{GuoRSZ2026} lossless-rank-condensers}\label{sec:cons3}

Our final construction of subspace list-recoverable codes uses the lossless-rank condensers of Guo, Raj, Shangguan, and Zhang \cite{GuoRSZ2026} (using \Cref{def:condeneser-to-SLRC}):
\begin{theorem}[\cite{GuoRSZ2026}]\label{thm:grsz}
    For any $R \in (0,1),r\in\mathbb{N}$, and any non-prime field $\mathbb F_q$ with $q>\poly(r)$, there exists an infinite explicit family of $\F_q$-additive codes forming lossless rank condensers up to rank $r$ with $s=O(r)$ and rate $\Omega(1/r)$. In particular, this family of codes forms an explicit infinite family of $(r,0,\ell, \ell)$ subspace list-recoverable codes for all $\ell \in \mathbb N$. Additionally, the code can be computed in time $\poly(n,r,\log q)$.
\end{theorem}

\section{Explicit Subspace Evasive Sets}\label{sec:explicit-SES}

In this section we combine the local-to-global reduction (\Cref{thm:framework}), the local optimal subspace evasive sets (\Cref{thm:inner}), and various explicit subspace list-recoverable codes from \Cref{sec:slcc} to explicitly construct subspace evasive sets.

Our first construction considers the case in which the field size $q$ is constant with respect to $n$. In this case, we make use of \Cref{cor:ggh1} and \Cref{cor:ggh2} derived from \cite{GoyalGH2026}.

\begin{theorem}\label{thm:sesggh}
Fix a prime power $q$ and constants $r\in\Z^+,\eps\in(0,1)$. For all large enough $n$, we can construct $(r,L=O(\min(\frac{r^2}{\eps},\frac{r\log{(r/\eps)}}{\eps})))$ subspace evasive sets $A\subseteq \F^n_q$ with size $|A|\ge q^{(1-\eps)n}$. We can construct a membership oracle for $A$ with running time $O_{q,r,\eps}(1)\poly(n)$.
\end{theorem}
\begin{proof}
We split the proof into two cases.
\begin{itemize}
\item (Case A) $L=O(\frac{r^2}{\eps})$, and
\item (Case B) $L=O(\frac{r\log{(r/\eps)}}{\eps})$.
\end{itemize} In these two different cases we will use \Cref{cor:ggh1} and \Cref{cor:ggh2} respectively. We let $D$ denote the reciprocal of the target rate of the subspace list recoverable code, i.e., $R = \frac{\dim_{\F_q}(C)}{sM} \geq \frac{1}{D}$. We then set the different parameters as follows.

\begin{itemize}
\item(Case A) $\ell=\lceil\frac{16r^2}{\eps}\rceil,D=2r$, and
\item(Case B) $\ell=\lceil\frac{256r(1+\log{(r/\eps)})}{\eps}\rceil,D=2+2\log_2{\ell}$. In this case $D\leq 20(1+\log_{2}(r/\eps))$.
\end{itemize}

Let $t=\lceil\frac{4D}{\eps}\rceil,s=s_0t,b=\lfloor\frac{\eps s}{2D}\rfloor,d=\ell+1$ where $s_0=O_{q,r,\eps}(1)$ is the constant $s$ in the statement of \Cref{cor:ggh1} (Case A) or \Cref{cor:ggh2} (Case B). Then we check the parameter requirement of 
\Cref{thm:inner} and invoke it: Since $b\ge 4r$, we know $b-r\ge \frac{\eps s}{4D}$. Therefore \[\frac{r(s+b)}{b-r}\leq \frac{r(5s/4)}{\eps s/4D}=\frac{5rD}{\eps}.\]
Splitting by case, we then compute that
\begin{itemize}
\item (Case A) $\frac{r(s+b)}{b-r}\leq \frac{5rD}{\eps}\leq \frac{10r^2}{\eps}<\ell+1=d$, and
\item (Case B) $\frac{r(s+b)}{b-r}\leq \frac{5rD}{\eps}\leq \frac{100r(1+\log{(r/\eps)})}{\eps}<\ell+1=d$.
\end{itemize}
In both cases, \Cref{thm:inner} provides a function $f\colon \F^s_q\to \F^b_q$ so that $G^f$ is a $(r,\ell)$ subspace evasive set with construction time $O_{q,r,\eps}(1)$.

By invoking \Cref{cor:ggh1} (Case A) or \Cref{cor:ggh2} (Case B), we get an $(r, 0, \ell, 2\ell)$ subspace list-recoverable code $\hat{C} \subseteq (\F^{s_0}_q)^{Mt}$ with rate at least $R \ge 1/D$ and encoding time $O_{q,r,\eps}(1)\poly(M)$ for all large enough $M$. Let $C \subseteq (\F^s_q)^M$ be the $\F$-additive code where groups of $t$ consecutive coordinates of $\hat{C}$ are bundled together. By inspection, such a code is also an $(r, 0, \ell, 2\ell)$ subspace list-recoverable code, has rate at least $R \ge 1/D$, and encoding time $O_{q,r,\eps}(1)\poly(M)$ for all large enough $M$.

Let $M=\lfloor\frac{\eps n}{b}\rfloor$, we know that $R=\frac{1}{D}\ge\frac{b}{\eps s}$. Therefore, $\dim(C)\ge \frac{bM}{\eps}\ge n-\frac{b}{\eps}\ge (1-\eps)n+1$ for large enough $n$. Let $k=n-bM$ and use a subcode of $C\subseteq (\F^s_q)^M$ with dimension $k$. By \Cref{thm:framework} we can construct a function $g\colon \F_q^k\to \F_q^{bM}$ such that $G^g\subseteq \F^n_q$ is an $(r,L=2\ell)$ subspace evasive set with construction and membership oracle time $O_{q,r,\eps}(1)\poly(n)$.
\end{proof}

The second construction replaces the SLRCs based on \Cref{thm:ggh} with SLRCs based on \Cref{thm:gk}. This is a more general construction since it works for all polynomial field sizes $q\leq \poly(n)$. However, the running time becomes $n^{O(rL)}$, which can be less efficient compared to $O_{q,\eps,\ell}(1)\poly(n)$ running time of the previous construction when $q$ is a constant.

\begin{theorem}\label{thm:sesgk}
For any constants $r \in \mathbb{Z}_{>0}$,
$\varepsilon \in (0,1)$, and $C>0$,
for all sufficiently large $n$ and every prime power
$q \leq n^C$, there exists an explicit
$(r,L)$ subspace evasive set $A \subseteq \F_q^n$
with
\[
L = O\left(\min\left\{
\frac{r^2}{\varepsilon},
\frac{r\log(r/\varepsilon)}{\varepsilon}
\right\}\right)
\]
and size $|A| \geq q^{(1-\varepsilon)n}$.
We can construct a membership oracle for $A$ with running time $n^{O_C(rL)}$.
\end{theorem}
\begin{proof}[Proof sketch]
This proof is similar to that of \Cref{thm:sesggh}. We split the proof into two cases.
\begin{itemize}
\item (Case A) $d=O(\frac{r^2}{\eps})$, and
\item (Case B) $d=O(\frac{r\log{(r/\eps)}}{\eps})$.
\end{itemize}
In these two different cases we will use \Cref{cor:gk1} and \Cref{cor:gk2} respectively and set different parameters as follows.
\begin{itemize}
\item(Case A) $\ell=\lceil\frac{50r^2}{\eps}\rceil,D=8r$, and
\item(Case B) $\ell=\lceil\frac{256r(1+\log{(r/\eps)})}{\eps}\rceil,D=3+3\log_2{\ell}$ so that $D\leq 33(1+\log_{2}(r/\eps))$.
\end{itemize}

Let $e$ denote the smallest positive integer such that $Q=q^e>8Dn$. We know $Q=\poly(n)$. Since an $(r,d)$ subspace evasive set $A\subseteq \F^{n/e}_Q$ with size at least $Q^{(1-\eps)n/e}$ actually implies an $(r,d)$ subspace evasive set $A\subseteq \F^n_q$ with size at least $q^{(1-\eps)n}$ via a linear bijection between $\F^e_q$ and $\F_Q$, it suffices to construct a subspace evasive set with the same quality but over the field extension $\F_Q$. Therefore, we can assume $q>8Dn$ without loss of generality.

Let $s_0=2r,t=\lceil\frac{4D}{\eps}\rceil,s=s_0t,b=\lfloor\frac{\eps s}{2D}\rfloor,d=\ell+1$. We follow analogous reasoning as in the proof of \Cref{thm:sesggh}. We only need to clarify the running time and check the parameter requirement of \Cref{cor:gk1} (Case A) or \Cref{cor:gk2} (Case B). 
Since $b\ge 4r$ and $M\leq \frac{\eps n}{b}$, it follows that $q>8Dn\ge 4rMt$, so we can indeed invoke \Cref{cor:gk1} (Case A) or \Cref{cor:gk2} (Case B).

The construction time from \Cref{thm:inner} is $q^{O((b+s+d)r)}=q^{O(rd)}=n^{O(rL)}$ since $L=2(d-1),d=\Omega(s)$ and $s\ge b$. Since the encoding time guaranteed by \Cref{cor:gk1} and \Cref{cor:gk2} is $\poly(n)$, by \Cref{thm:framework} is final construction time and membership oracle time can be bounded within $n^{O(rL)}$.
\end{proof}
The third construction uses \cite{GuoRSZ2026} based SLRCs instead. It provides better construction time in many cases. However, it only achieves intersection size $O(r^2/\eps)$.
\begin{theorem}\label{thm:sesgrsz}
Fix $r\in\Z^+,\eps\in(0,1)$ and prime power $q$, for infinitely many $n$, we can construct $(r,O(r^2/\eps))$ subspace evasive sets $A\subseteq \F^n_q$ with  size $|A|\ge q^{(1-\eps)n}$. We can construct a membership oracle for $A$ with running time $\max(r,q)^{O(r^3/\eps)}+\poly(n,r,\log{q})$.
\end{theorem}
\begin{proof}[Proof sketch]
The proof is basically the same as the (Case A) of the above proof of \Cref{thm:sesgk}. The only two differences are: (i) We replace \Cref{cor:gk1} with \Cref{thm:grsz}, with an advantage that we only need to assume $q>\poly(r)$ rather than $q>8Dn$; and  (ii) Some parameter settings need to be scaled by constant multiplicative factors to match the statement of \Cref{thm:grsz}. 
\end{proof}

\section{Explicit List-recoverable Codes}\label{sec:explicit-LR}

In this section we present four explicit constructions of $(1-R-\eps,\ell,L=\poly(\ell/\eps))$ list-recoverable codes. All of these results share are based on the paradigm introduced by Guruswami and Wang~\cite{GuruswamiW2013} of taking a subcode of an additive code using a subspace evasive set. We begin by defining a notion of \emph{dimension} list recovery.

\begin{definition}\label{def:DLR}
For integers $\ell, r \ge 1$ and $\rho \in [0, 1]$, we say that an additive code $C \subseteq (\F_q^s)^n$ is $(\rho, \ell, r)$ \emph{dimension} list-recoverable if for any sets $S_1, \hdots, S_n \in \binom{\F_q^s}{\le \ell}$, the list
\[
\mathcal L = \{c \in C  \,\colon\, \exists y \in (S_1 \times \cdots \times S_n), \Ham(c, y) \le \rho n\}.
\]
lies in an $r$-dimensional subspace of $(\F_q^s)^n$.
\end{definition}

Unlike our earlier notion of subspace list recovery (Defintion~\ref{def:slrc}) which intersects the recovered list $\mathcal L$ with a low-dimensional subspace, Definition~\ref{def:DLR} says that $\mathcal L$ \emph{always} lies in a low-dimensional subspace. Despite being a rather strict notion, all the families of additive codes we consider in this paper, when invoked with suitable parameters, are $(1-R-\eps,\ell,\poly(\ell/\eps))$ dimension list-recoverable. If we assume our additive code has an encoding map $\Enc : \F_q^k \to (\F_q^s)^n$, by plugging in a suitable subspace evasive set $A \subseteq \F_q^k$ into the encoder, the \emph{subcode} $\Enc(A) \subseteq C$ has the property that its recovered list $\mathcal L$ lies in a low-dimensional subspace \emph{and} this subspace has few elements in $\Enc(A)$. We formalize this observation as follows.


\begin{theorem}\label{thm:SES-subcode}
For prime power $q$, integers $s, n, k, r, c,\ell \ge 1$ and $R, \eps, \rho > 0$. Let $A \subseteq \F_q^k$ be an $(r, c)$ subspace evasive set of size at least $q^{(1-\eps/2)k}$ that is constructable in time $T_A$. Let $\Enc : \F_q^k \to (\F_q^{s})^n$ be the encoding map of a code $C$ of rate $\frac{k}{sn} \ge R+\eps$ which is $(\rho, \ell, r)$ dimension list-recoverable. Then $\Enc(A) \subseteq (\F_q^s)^n$ is an $(\rho, \ell, c)$ list-recoverable code of rate at least $R$ with encoding time $T_A + T_{\Enc}$, where $T_{\Enc}$ is the runtime of $\Enc$.
\end{theorem}
\begin{proof}
For lists $S_1, \hdots, S_n \subseteq \F_q^s$ of size at most $\ell$, we construct
\[
\mathcal L = \{c \in C  \,\colon\, \exists y \in (S_1 \times \cdots \times S_n), \Ham(c, y) \le \rho n\}.
\]
Since $C$ is $(\rho, \ell, r)$ dimension list-recoverable, we have that $\rank \mathcal L \le r$. Since $A$ is an $(r, c)$ subspace evasive set, we have that $A \cap \Enc^{-1}(\mathcal L)$ has at most $c$ elements. Thus,
\[
\{c \in \Enc(A)  \,\colon\, \exists y \in (S_1 \times \cdots \times S_n), \Ham(c, y) \le \rho n\} \subseteq \Enc(A) \cap \mathcal L = \Enc(A \cap \Enc^{-1}(\mathcal L))
\]
has at most $c$ elements. Thus, $\Enc(A)$ is $(\rho, \ell, c)$ list-recoverable. Furthermore, the rate of $\Enc(A)$ is at least $\frac{(1-\eps/2) k}{sn} \ge (1 - \eps/2)(R+\eps) \ge R$. The encoding time of $T_A + T_{\Enc}$ follows from the construction being a direct composition.
\end{proof}

By picking various additive codes (RS, AEL, AG, and FRS) and invoking \Cref{thm:SES-subcode}, we achieve a rich family of constructions of near-optimal list-recoverable codes. We now formally describe each of these constructions.

\subsection{Subcodes of Reed--Solomon Codes}
We first construct explicit subcodes of Reed--Solomon Codes with good list-recoverability. It builds upon the following additive subcodes of RS codes
\begin{theorem}[\cite{gz26}, list-recovery generalization of \cite{gx22}---see also {\cite[Section 5.3]{ldsurvey}}]\label{thm:subrs-pre}
For any constants $\ell\ge 1,\eps,R\in(0,1)$, prime power $q\ge n$ and distinct evaluation points $\alpha_1,\dots,\alpha_n\in\F_q$, there exists $s=O(1/\eps^2)$ and explicit $\F_{q}$-additive Reed-Solomon subcode $C \subseteq (\F_{q^s})^n$ of rate $R$ such that $C$ is $(1-R-\eps, \ell, O(\ell/\eps^2))$ dimension list-recoverable.  
\end{theorem}

Composing \Cref{thm:sesgk}, \Cref{thm:subrs-pre} (with $\eps = \eps/2)$, and \Cref{thm:SES-subcode}, we get the following explicit list-recoverable code.


\begin{theorem}\label{thm:subrs-main-res}
For any constants $\ell \geq 1$, $\varepsilon \in (0,1)$,
$R \in (0,1-\varepsilon)$, and $M \geq 1$,
for all sufficiently large $n$, every prime power
$n \leq q \leq n^M$, and distinct evaluation points
$\alpha_1,\ldots,\alpha_n \in \F_q$,
there exists $s=O(1/\varepsilon^2)$ and an explicit
subcode of a Reed--Solomon code over $\F_{q^s}$
with evaluation points $\alpha_1,\ldots,\alpha_n$
and rate at least $R$ that is
\[
\left(1-R-\varepsilon,\ell,
O\left(\frac{\ell}{\varepsilon^3}
\log\frac{\ell}{\varepsilon}\right)\right)
\]
list-recoverable. The encoding time is
$n^{\poly_M(\ell/\varepsilon)}$.
\end{theorem}

\begin{remark}\label{remark:AGG}
Very recently, Alrabiah, Goyal, and Guruswami~\cite{AlrabiahGG26a} constructed explicit Reed--Solomon codes which (among other properties) are $(1-R-\eps, \ell, O(\ell/\eps))$ dimension list-recoverable. By composing that result with \Cref{thm:sesgk} and \Cref{thm:SES-subcode} one can construct codes of rate at least $R$ that are $\left(1-R-\varepsilon,\ell,O\left(\frac{\ell}{\varepsilon^2}\log\frac{\ell}{\varepsilon}\right)\right)$ list-recoverable, approximately the same as our \Cref{thm:subfrs-main-res}. 
\end{remark}

\subsection{Subcodes of AEL Codes}
Our next construction of an explicit list-recoverable codes takes advantage of identifying a subcode of AEL-based subspace design codes in \cite{GoyalGH2026}.

\begin{theorem}\label{thm:ael-pre}
For any constants $\ell\ge 1,\eps,R\in(0,1)$ and prime power $q$, for $s=\exp(O(\ell^2/\eps^2)\log{q})$, there exist explicit $\F_{q}$-additive codes over alphabet $\F^s_{q}$ with rate at least $R$ which are $(1-R-\eps, \ell, O(\ell/\eps))$ dimension list-recoverable.
\end{theorem}
\begin{proof}
Let $\tau(r)=R+\eps/2$ for all $r\leq \frac{2\ell}{\eps}$ and $\tau(r)=1$ for $r>\frac{2\ell}{\eps}$. By \cite[Theorem 1.1]{GoyalGH2026}, there exist explicit $\tau$-subspace design codes with the prescribed rate, $\F_q$-additivity, and alphabet size. Then, by \cite[Lemma 2.10]{GoyalG2025-Structure}, such codes satisfies the required property that all messages concerned in the statement lie within a certain $\frac{2\ell}{\eps}$-dimensional subspace of $\F^k_q$.
\end{proof}

Composing \Cref{thm:sesggh}, \Cref{thm:ael-pre} (with $\eps = \eps/2)$, and \Cref{thm:SES-subcode}, we get our next explicit list-recoverable code.

\begin{theorem}\label{thm:subael-main-res}
For any constants $\ell\ge 1,\eps,R\in(0,1)$, there are explicit codes with rate at least $R$, alphabet size $\exp(\exp(O(\ell^2/\eps^2)))$ that are $(1-R-\eps,\ell,O(\frac{\ell}{\eps^2}\log{\frac{\ell}{\eps}}))$ list-recoverable. The encoding time is $O_{\ell,\eps}(1)\poly(n)$.
\end{theorem}

\subsection{Subcodes of AG Codes}
Similarly, we also have explicit linear subcodes of AG codes with good list-recoverability. This is due to a generalization of \cite{grz21} from list-decoding to list-recovery.

\begin{theorem}[\cite{gz26}, list-recovery generalization of \cite{grz21}]\label{thm:subag-pre}
For any constants $\ell\ge 1,\eps,R\in(0,1)$, there exists a prime power $q=O(\frac{1}{\eps^2})$, an extension $\F_{q^m}$ with size $q^m=(\frac{\ell}{\eps})^{O(\ell/\eps^2)}$ and explicit  $\F_q$-linear subcodes of AG codes over $\F_{q^m}$ with rate $R$ which are $(1-R-\eps, \ell, \poly(\ell/\eps))$ dimension list-recoverable.
\end{theorem}


We next compose \Cref{thm:sesggh}, \Cref{thm:subag-pre} (with $\eps = \eps/2)$, and \Cref{thm:SES-subcode} to get the following construction.

\begin{theorem}\label{thm:subag-main}
For any constants $\ell\ge 1, 0<\eps<1$ and $R \in (0, 1-\eps)$, there are explicit subcodes of algebraic-geometry codes over $\F_{q}, q=(\frac{\ell}{\eps})^{O(\ell/\eps^2)}$ with rate at least $R$ that are $(1-R-\eps,\ell,\poly(\ell/\eps))$ list recoverable. The encoding time is $O_{\ell,\eps}(1)\poly(n)$.
\end{theorem}

Although the list size here is worse than before (with a exponent that is not explicitly calculated), it has the smallest-known alphabet size $\exp(\poly(\ell/\eps))$ of an explicit $(1-R-\eps,\ell,\poly(\ell/\eps))$ list-recoverable code. Even in the classical list-decoding literature where $\ell=1$, all previous explicit $(1-R-\eps,\poly(1/\eps))$ list-decodable codes require an alphabet of size at least $\exp(\exp(\poly(1/\eps)))$ (and which is achieved by \cite{JeronimoMST2025}).

\subsection{Subcodes of Folded Reed--Solomon Codes}
Finally we give explicit list-recoverable codes through a subspace evasive subcode of the folded Reed--Solomon  codes in \cite{GuruswamiR2008}. Such subcodes of folded RS codes were in fact the motivating use of subspace evasive sets in the list decoding context in \cite{guru11,GuruswamiW2013}.

\begin{theorem}[{\cite[Theorem 7]{GuruswamiW2013}}]\label{thm:frs-pre}
For any constants $\ell\ge 1,\eps,R\in(0,1),s\ge 16\ell/\eps^2$ and $q\ge sn$, there exist  explicit $\F_{q}$-folded Reed--Solomon codes over $\F^s_{q}$ with rate at least $R$ such that the code is $(1-R-\eps, \ell, O(\ell/\eps))$ dimension list-recoverable.
\end{theorem}

To finish, we compose \Cref{thm:sesgk}, \Cref{thm:frs-pre} (with $\eps = \eps/2)$, and \Cref{thm:SES-subcode} to get the following construction.

\begin{theorem}\label{thm:subfrs-main-res}
For any constants $\ell\ge 1,\eps,R\in(0,1)$, there are explicit subcodes of folded Reed--Solomon codes with rate at least $R$, alphabet size $(\frac{n\ell}{\eps^2})^{O(\ell/\eps^2)}$ that are $(1-R-\eps,\ell,O(\frac{\ell}{\eps^2}\log{\frac{\ell}{\eps}}))$ list-recoverable. The encoding time is $n^{\poly(\ell/\eps)}$.
\end{theorem}

Despite the large alphabet, this construction yields efficient list-recovery algorithms thanks to recent algorithmic advances \cite{GoyalG2025-Structure,ghks26}. This is explained in detail as in \Cref{sec:frs-alg}.

\section{Algorithmic subspace evasion and list-recovery}\label{sec:algorithm}


We show that any of the above subspace evasive constructions can moreover be made algorithmic, i.e., given a low-dimensional input subspace, its intersection with the subspace evasive set can be computed efficiently. We then use this to make the list-recovery claim for the codes of~\Cref{thm:subfrs-main-res} algorithmic.

\subsection{Algorithmic version of subspace evasive set construction}
The following result adds efficient intersection computation to the subspace evasive set construction of~\Cref{thm:framework}
\begin{theorem}
\label{thm:algorithmic-evasion}
    Let $\Enc_1, \dots, \Enc_n: \F_q^k \mapsto \F_q^s$ be linear maps constructible in time $T_C$ such that the resulting code $C = \{(\Enc_1(x), \dots, \Enc_n(x)),x\in\F^k_q\}\subseteq (\F_q^s)^n$ is an $r$-lossless rank condenser, and let $f:\F_q^s \mapsto \F_q^b$ be so that $G^f$ is an $(r, c_f)$ subspace-evasive set. Assume that given any linear space $A\subseteq \F_q^{s+b}$ of dimension $\leq r$, $G^f\cap A$ can be computed in time $\mathsf{Dec}_f$. 

    Then, we can construct a function $g: \F_q^k \mapsto \F_q^{bn}$ such that $G^g$ is an $(r, c_f)$ subspace-evasive set of size $q^k$, and $G^g$ is constructable, and moreover given any subspace $U\subseteq \F_q^{k+bn}$ of dimension $\leq r$, $U\cap G^g$ can be computed in time $\mathsf{Dec}_f+(c_f+T_C)\cdot\poly(n, s, \log q)$
\end{theorem}
\begin{proof}
    We let $g(x)= (f(\Enc_1(x)), \dots, f(\Enc_n(x)))$ as in the proof of \Cref{thm:framework}. Observe that by \Cref{def:condeneser-to-SLRC}, the condenser is an $(r, 0, c_f, c_f)$ subspace list-recoverable code. Thus, by \Cref{thm:framework}, the resulting construction is subspace-evasive. We now simply need to show that it is also ``decodable" efficiently.

    Let the given linear space be $U\subseteq \F_q^{k+bn}$ of dimension $\leq r$. Now, let $U'$ be the projection of $U$ onto the first $k$ coordinates. Let $\dim U'=r'\leq r$. Now, by \Cref{def:lossless-rank-condenser}, there is a coordinate $i\in [n]$ such that $\dim(\Enc_i(U')) = r'$. Observe that such an $i$ can be computed in time $T_C\poly(n, s, r, \log q)$ by simply computing $\dim (\Enc_i(U'))$ for all $i\in [n]$. 
    
    Now, for all $u\in U$, let $u'$ be the projection of it onto the first $k$ coordinates and $u'_i$ be the projection onto the following $i$th block of $b$ coordinates. Observe that if $u \in U\cap G^g$, we must have that $u_i = f(\Enc_i(u'))$. That is, $(\Enc_i(u'), u_i)\in G^f$. Thus, we can now simply compute $\{(\Enc_i(u'), u_i)\,\colon\, u'\in U'\}\cap G^f$ in $\mathsf{Dec}_f+T_C\cdot \poly(n, \log q, s)$ time.

    Since $\Enc_i$ is injective on $U'$, we get that we can thus recover $u'$ for all $u'\in U'$ such that there is an $u\in U\cap G^g$  
    with $u_{\leq k}=u'$. Additionally, there are at most $c_f$ such choices of $u'$.

    Now, it simply suffices to check which of them are in $G^g \cap U$ which takes $c_f \cdot T_C\cdot\poly(n, s, \log q) $ time.
\end{proof}

\begin{corollary}\label{cor:algorithmic-evasion}
    For any $\eps\in (0,1)$, for large enough $k$ and for any $(r, c_A)$ subspace-evasive set $A\subseteq (\F_q^k)$ that is constructable and has membership oracle in time $T_A$, there is an $A' \subseteq \F_q^{k+\lfloor\eps k\rfloor}$ which is also $(r, c_A)$ subspace-evasive with $|A'|=|A|$, constructible in time $O_{q,r,\eps}(1)\cdot \poly(k)+c_AT_A$, and for any $U\subseteq \F_q^{k+\lfloor\eps k\rfloor}$ of dimension $\leq r$, $A'\cap U$ and can be computed in time $O_{q, r,\eps}(1)\poly(k)+c_AT_A$ time.
\end{corollary}
\begin{proof}
    Observe that in \Cref{thm:algorithmic-evasion}, we can pick the lossless rank-condenser to be of $s=O_{q, r, \eps}(1)$ dimension and $f$ from using \Cref{thm:inner} to be such that $bn \leq \eps k$ such that the resulting set $G^g$ is $(r, \poly(r, 1/\eps))$ subspace-evasive using \Cref{sec:explicit-SES}.

    Now, let $A' = \{(a, g(a))\,\colon\, a\in A\}$. Observe that $|A'|=|A|$ and $A'\cap U\subseteq G^g \cap U$ which can be computed in time $O_{q, r,\eps}(1)\poly(k)$. But, now restricting to the first $k$ coordinates, $A$ is already $(r, c_A)$ subspace-evasive. Thus, we can now check membership in $A$ for all elements of $(G^g \cap U)_{\leq k}$ with an additional $c_AT_A$ time.
\end{proof}

\subsection{Consequences for list-recovery}\label{sec:frs-alg}

One key drawback of the above algorithmic results has been that the runtime scales as $n^{\poly(\ell/\eps)}$ when $q\ge n$. To see this, we observe that all of \Cref{thm:sesggh,thm:sesgk,thm:sesgrsz} require $\mathsf{Dec}_f=q^{\poly(\ell/\eps)}$. Moreover, in almost all algorithmically list-recoverable codes with constant dimension, the corresponding alphabet size $q$ is larger than $n$, and thus the existence of an $O_{\ell,\eps}(1)n^{O(1)}$ complexity list-decoder is unclear. We provide a method to sidestep this obstruction by constructing subcodes of folded Reed--Solomon codes that are $(1-R-\eps,\ell,\poly(\ell/\eps))$ list-recoverable with a randomized list-recovery algorithm running in $O_{R,\ell,\eps}(1)n^{O(1)}$ time.
\begin{remark}
    We note that there are algorithmically $(1-R-\eps, \ell, O_{\ell, \eps}(1))$ list-recoverable constructions of rate $R$ in the literature with alphabet $O_{\ell, \eps, R}(1)$ as well, but as far as we know there is no stated result showing that these constructions have their list contained in a $\poly(\ell/\eps)$-dimensional space. The construction in \cite{GoyalGH2026} does have the second feature, however no list recovery algorithm was presented. We suspect that this should be possible by employing the techniques of \cite{JeronimoMST2025, SrivastavaT2025}, but as far as we know, this has not been established in the literature. Here, we present an alternative method, of possible independent interest, that circumvents this gap.
\end{remark}

We remind ourselves of a result from \cite{GoyalG2025-Structure} about the list-recovery of the folded Reed-Solomon codes of \cite{GuruswamiR2008}. To state their result, we require the definition of a \emph{sum set}.

\begin{definition}[Sum Sets]
    $P\subseteq \mathcal C$ is an $(u, v)$ \textit{sum-set} in $\mathcal C$ if it is described by sets $A_1, A_2, \cdots, A_{u'}$ with $A_i\subseteq \mathcal C$, $|A_i|\le v$, and $u'\le u$ such that $P=A_1+\cdots +A_{u'}=\{a_1+a_2+\ldots + a_{u'} \,\colon\, a_i \in A_i\}$.
\end{definition}
\begin{theorem}
\label{thm:FRS-nearlinear}
    For every choice of $\ell\in \mathbb N, \eps>0, R>0$, any rate $R$ $s$-folded Reed--Solomon code $\mathcal C$ for $s>s_0 := \frac{16(R+\eps)\ell}{\eps^2}$ has the following property.\footnote{Folded Reed--Solomon codes can be constructed for any $q\ge sn$ and do not require characteristic assumptions.}

    There exists an algorithm that takes in input lists $L_1, \cdots, L_n \subseteq \F_q^s$ of size $\ell$ each, runs in time $\poly(n,s,\log q)$, and outputs the descriptions of $t = \widetilde O(\ell^2/\eps^3)$ many
    \[
      \left(\left\lceil \frac{2(R+\eps)\log(2e\ell/(R+\eps))}{\eps}\right\rceil,\ell\right)
      \text{ sum-sets } P_1,\cdots,P_t
    \]
    such that
    $\{c\in\mathcal C\,\colon\, \Delta(c,L_1\times\cdots\times L_n)<1-R-\eps\}\subseteq\bigcup_{i=1}^t P_i$
    with probability $\ge\frac12$.
\end{theorem}

Finally, we make use of the above results to achieve a $O_{R,\ell,\eps}(1)n^{O(1)}$ time list-recovery algorithm. Broadly speaking, since we can efficiently find several sum-sets in a folded RS code and all concerned codewords are in their union, it suffices to compute intersections of the linear spans of these sum-sets with a subspace evasive set. Each sum-set is contained in an $\F_2$-linear subspace of dimension at most $u\ell$, so it is already enough to have subspace evasive sets over $\F_2$ rather than $\F_q$ when $q$ has characteristic $2$, and subspace evasive sets over $\F_2$ yield efficient intersection-solving algorithms.

\begin{theorem}\label{thm:alg-frs-main}
    For any $\eps\in(0,1)$, $\ell\in\N$, and rate $R\in(0,1-\eps)$, there is a constant $s=O(\ell/\eps^2)$, and an explicit family of codes $C\subseteq(\F_q^s)^n$ of rate $\ge R$, for all sufficiently large $n$ and a choice of $q=2^d$ with $sn\le q<2sn$, constructible in time $O_{R,\ell,\eps}(1)\poly(n)$ and $(1-R-\eps,\ell,\poly(\ell,1/\eps))$ list-recoverable.

    Moreover, there is a randomized algorithm given any $S_1,\cdots,S_n\subseteq\F_q^s$ with $|S_i|\le\ell$ for all $i\in[n]$, such that $\{c\in C\,\colon\,\Delta(c,S_1\times\cdots\times S_n)\le1-R-\eps\}$ can be output in time $O_{\eps,\ell}(1)\poly(n)$ with probability at least $1/2$.
\end{theorem}

\begin{proof}
Pick $s=\lceil64\ell/\eps^2\rceil$ and $q_0=2$, and let $d$ be the smallest positive integer such that $2^d\ge sn$. Let $\F_q$ be the field of size $q=2^d$ and let $k=\lceil(R+\eps/2)sn\rceil-1$ and $k_0=dk$.

    We let $r=\lceil 10\ell\log(2e\ell/\eps)/\eps\rceil$ and $\eta=\eps/8$, and set $m=\lfloor k_0/(1+\eta)\rfloor$. By \Cref{thm:sesggh}, we choose $A_0\subseteq\F_{q_0}^m$ to be an $(r,\poly(\ell,1/\eps))$ subspace-evasive set with $|A_0|\ge2^{(1-\eta)m}$. By \Cref{cor:algorithmic-evasion}, this gives a subspace-evasive set in $\F_{q_0}^{m+\lfloor\eta m\rfloor}$ with the same size and intersection bound, whose intersection with any subspace of dimension at most $r$ can be computed in time $O_{\eps,\ell}(1)\poly(m)$. Padding with zero coordinates, we let $A\subseteq\F_{q_0}^{k_0}$ be the resulting subspace-evasive set. We have $|A|\ge2^{(1-\eps/2)k_0}$ for all sufficiently large $n$, and given any $U\subseteq\F_{q_0}^{k_0}$ of dimension at most $r$, $A\cap U$ can be output in time $O_{\eps,\ell}(1)\poly(k_0)=O_{\eps,\ell}(1)\poly(n)$. Going forward, for clarity in the argument, we let $L:\F_{q_0}^{k_0}\to\F_q^k$ be an $\F_{q_0}$-linear isomorphism.

    Now, let $C_0$ be the $s$-folded Reed--Solomon code of dimension $k$ over $\F_q$ and block-length $n$. Let it be given by $\{\Enc(x)\,\colon\, x\in\F_q^k\}$.

    Let $C=\{\Enc(L(a))\,\colon\, a\in A\}$. Observe that $|C|=|A|$. Thus, the resulting code has rate
    \[
      \frac{\log_2|A|}{dsn}\ge(1-\eps/2)\frac{k}{sn}
      \ge(1-\eps/2)\left(R+\frac\eps2-\frac1{sn}\right)\ge R,
    \]
    for all sufficiently large $n$, where the last inequality uses $R<1-\eps$.

    We now show the required list-recovery guarantee. Fix $S_1,\ldots,S_n\subseteq\F_q^s$ of size at most $\ell$ each, and let $R_0=k/(sn)$. Observe that by \Cref{thm:FRS-nearlinear}, applied to $C_0$ with rate $R_0$ and error parameter $\eps/2$, there is an algorithm that outputs $P_1,\dots,P_t$ which are $(u,\ell)$ sum-sets in $C_0$ in time $O_{\eps,\ell}(1)\poly(n)$ with $t=\widetilde O(\ell^2/\eps^3)$, where
    \[
       u=\left\lceil\frac{4(R_0+\eps/2)\log(2e\ell/(R_0+\eps/2))}{\eps}\right\rceil,
       \qquad u\ell\le r,
    \]
    such that
    \[
       \{c\in C\,\colon\,\Delta(c,S_1\times\cdots\times S_n)\le1-R-\eps\}
       \subseteq\bigcup_{i=1}^t P_i
    \]
    with probability at least $1/2$. Here, we pad the input sets to size $\ell$ if necessary. The theorem applies since $s>64(R_0+\eps/2)\ell/\eps^2$ and $R_0< R+\eps/2$, so its guaranteed radius is larger than $1-R-\eps$.

    Now, we can observe that, for each $i$, we can write $P_i=A_{i,1}+\cdots+A_{i,u_i}$ with $u_i\le u$ and $|A_{i,j}|\le\ell$, and let
    \[
       V_i=\Span_{\F_{q_0}}\left(\bigcup_{j=1}^{u_i}A_{i,j}\right)\subseteq C_0.
    \]
    Since $P_i\subseteq V_i$ and $\dim_{\F_{q_0}}(V_i)\le u\ell\le r$, we can compute
    \[
       A\cap L^{-1}(\Enc^{-1}(V_i))
    \]
    in time $O_{\ell,\eps}(1)\poly(n)$, and it has size $\poly(\ell,1/\eps)$ due to the subspace-evasive property of $A$. The preimage is an $\F_{q_0}$-linear subspace, and a basis for it can be computed efficiently from the sum-set description.

    Finally, observe that since $C\subseteq C_0$, we can compute $C\cap\left(\bigcup_{i=1}^t V_i\right)$ by encoding the points in these intersections, and retain exactly those codewords satisfying $\Delta(c,S_1\times\cdots\times S_n)\le1-R-\eps$. Since $P_i\subseteq V_i$, this outputs the required list in time $O_{\ell,\eps}(1)\poly(n)$ with probability at least $1/2$. Moreover, the list size is at most $t\cdot\poly(\ell,1/\eps)=\poly(\ell,1/\eps)$.
\end{proof}

\section*{AI Methodology.}
The results in this paper were mostly developed in May and June 2026 without the use of LLMs. We prompted GPT models 5.5/5.6 with our constructions, but they could not improve the stated bounds. 
Some LLM assistance was used to improve the overall writing quality. 

Recently we fed a draft of the paper into GPT-6 and it was able to improve the bound $c$ in a $(r,c)$-subspace evasive set construction to the optimal $O(r/\epsilon)$ bound. This is via an entirely different approach. We will report this separately after understanding and rewriting the construction and proof. We note that our $\tilde{O}(r/\eps)$ bound is almost optimal and already an exponential improvement over the previous bound. 

\bibliography{main, coding-theory}

\appendix

\section{Comprehensive Table of Prior Work}

In \Cref{tab:listrecovery-app}, we present a comprehensive version of \Cref{tab:listrecovery} which details essentially all prior works on list-recovery.

\begin{center}
\begin{table}
\footnotesize
\begin{tabular}{|m{7cm}|m{3cm}|m{3cm}|m{2cm}|m{1.4cm}|}
\hline 
Codes & Alphabet size $|\Sigma|$ & List size $L$ & Explicit?    \\ \hline

Random Linear Codes \cite{ZP81,guruthesis,rw18,ls25} & $\ell^{O(1/\eps)}$ & $(\ell/\eps)^{O(\ell/\eps)}$ & No\\
\hline
Random Linear Codes \cite{BrakensiekCDZ2025-BrascampLieb} &$\ell^{(\ell/(R+\eps))^{O(1+R/\eps)}/\eps}$ & $\left(\frac{\ell}{R+\epsilon}\right)^{O(1+R/\eps)}$&No \\ 
 \hline
 Random Linear Codes \cite{gg26b} & $(\frac{\ell}{R+\eps})^{O((R+\eps)/\eps^2)}$ & $\left(\frac{\ell}{R+\epsilon}\right)^{O(1+R/\eps)}$ & No\\
\hline
Random Codes & $\ell^{O(1/\eps)}$ & $O(\ell/\eps)$ & No\\
\hline
 \hline 

Folded RS Codes \cite{GuruswamiR2008} & $\left( \frac{n}{\eps^2} \right)^{O(\log(\ell)/\eps^2)}$& $\left( \frac{n}{\eps^2} \right)^{O(\log(\ell)/\eps^2)}$ & Yes    \\ \hline 
Folded RS Codes \& Subspace Evasive Sets \cite{dl12} & $\left( \frac{n}{\eps^2} \right)^{O(\log(\ell)/\eps^2)}$& $(\frac{1}{\eps})^{O(1/\eps^3)}$ & Yes    \\ \hline 
Univariate Multiplicity Codes \cite{kopparty2015list} & $\left( \frac{n}{\eps^2} \right)^{O(\log(\ell)/\eps^2)}$& $\left( \frac{n}{\eps^2} \right)^{O(\log(\ell)/\eps^2)}$ & Yes   \\ \hline 
Folded RS/Univariate Multiplicity Codes \cite{GuruswamiW2013} & $\left( \frac{n \ell}{\eps^2} \right)^{O(\ell/\eps^2)}$ & $\left( \frac{n \ell}{\eps} \right)^{O(\ell/\eps)}$ & Yes   \\ \hline
Folded RS/Univariate Multiplicity Codes \cite{kopparty2018improved, tamo24} & $\left( \frac{ n \ell}{\eps^2} \right)^{O(\ell/\eps^2)}$& $(\ell/\eps)^{O((1+\log{\ell})/\eps)}$ & Yes     \\ \hline 
Folded RS plus AEL Codes \cite{kopparty2018improved} & $(\frac{\ell}{\eps})^{O(\ell^2\log{(\ell/\eps)}/\eps^2)}$& $\ell^{O(1/\eps^4)}$ & Yes     \\ \hline 
Folded AG Codes \cite{GuruswamiX2013, GuruswamiK2016} & $\exp\left( \frac{ \ell \log(\ell/\eps) }{\eps^2} \right)$ & $2^{2^{2^{O_{\eps,\ell}(\log^*(n))}}}$ & Yes  \\ \hline
AG Codes + BTT evasive subspace \cite{grz21} & $(\frac{\ell}{\eps})^{O(\ell/\eps^2)}$ & $\exp(\poly(\ell/\eps))$ & Yes  \\ \hline
Codes Based on the AEL Framework \cite{SrivastavaT2025}& $\exp(\exp(\ell/\eps\log{(\ell/\eps)}))$ & $ \exp(\exp(\ell/\eps\log{(\ell/\eps)}))$ &Yes \\\hline

Codes Based on the AEL Framework \cite{JeronimoS2025}& $\exp((\ell/\eps)^{(\ell/\eps)^{(\ell/\eps)}})$ & $(\ell/\eps)^{O(\ell/\eps)}$ &Yes \\\hline

Folded RS/Univariate Multiplicity Codes \cite{BrakensiekCDZ2025-BrascampLieb}& $\left( \frac{ n \ell}{\eps^2} \right)^{O(\ell/\eps^2)}$ &$\left(\frac{\ell}{R+\epsilon}\right)^{O(1+R/\eps)}$&Yes\\
\hline
Codes Based on the AEL Framework \cite{GoyalGH2026}& $\exp(\exp(O(\ell^2/\eps^2)))$ & $\left(\frac{\ell}{R+\epsilon}\right)^{O(1+R/\eps)}$ &Yes \\
\hline
RS codes + Subspace Evasive Sets \textbf{(\Cref{thm:subrs-main-res})}& $n^{O(1/\eps^2)}$ & $O(\frac{\ell}{\eps^3}\log{\frac{\ell}{\eps}})$ &Yes \\
\hline
AG codes + Subspace Evasive Sets \textbf{(\Cref{thm:subag-main})}& $(\frac{\ell}{\eps})^{O(\ell/\eps^2)}$ & $\poly(\ell/\eps)$ &Yes \\
\hline
Folded RS codes + Subspace Evasive Sets \textbf{(\Cref{thm:subfrs-main-res})} & $(\frac{n\ell}{\eps^2})^{O(\ell/\eps^2)}$ & $O(\frac{\ell}{\eps^2}\log{\frac{\ell}{\eps}})$ &Yes \\
\hline
AEL codes + Subspace Evasive Sets \textbf{(\Cref{thm:subael-main-res})}& $\exp(\exp(\ell^2/\eps^2))$ & $O(\frac{\ell}{\eps^2}\log{\frac{\ell}{\eps}})$ &Yes \\
 \hline
 \hline
 Any Linear Code \cite{guru21,ChenZ2025,LeviMS2025,ls25} & Any & $\ell^{\Omega(R/\eps)}$ & Lower Bound\\
 \hline
 Classical AEL Codes \cite{bcpz26} & Any & $\ell^{\Omega(R/\eps)}$ & Lower Bound\\
 \hline
\end{tabular}
\caption{This table gives a summary of many constructions of list-recoverable codes as well as some known limitations.} \label{tab:listrecovery-app}
\end{table}
\end{center}
\end{document}